\documentclass{pinchcr} 

\usepackage{graphicx}
\usepackage{amsmath}

\title{An overview of the theories of the glass transition}

\author{Gilles Tarjus}
\affiliation{LPTMC,  CNRS-UMR 7600, Universit\'e 
Pierre et Marie Curie, 4 Place Jussieu, 75252 Paris cedex 05, France}

\begin{document}

\maketitle

\preface

The topic of the glass transition gives rise to a a wide diversity of views. It is, accordingly, characterized by a lack of agreement on which would be the most profitable theoretical perspective. In this chapter, I provide some elements that can help sorting out the many theoretical approaches, understanding their foundations, as well as discussing their validity and mutual compatibility. Along the way, I describe the progress made in the last twenty years, including new insights concerning the spatial heterogeneity of the dynamics and the characteristic length scales associated with the glass transition. An emphasis is put on those theories that associate glass formation with growing collective behavior and emerging universality. 
\\

\noindent
\textbf{Contents}
\\

\noindent
1 Introduction
\\

\noindent
2 A diversity of views and approaches

2.1 What is there to be explained?

2.2 Sorting out the theoretical approaches
\\

\noindent
3 Elements of theoretical strategies

3.1 Minimal models and simplifying concepts

3.2 Looking for a localized relaxation mechanism in real space

3.3 Developing a specific statistical-mechanical framework

3.4 In search of a growing length scale
\\

\noindent
4 Theories based on an underlying dynamical transition

4.1 Mode-coupling theory

4.2 Dynamical facilitation and kinetically constrained models
\\

\noindent
5 Theories based on an underlying thermodynamic or static transition

5.1 Free-volume and  configurational-entropy models

5.2 Random first-order transition theory

5.3 Frustration-based approach

5.4 Jamming scenario
\\


\noindent
6 Concluding remarks
\\

\noindent
References
\\

\maintext

\section{Introduction}
\label{sec:1}

Anyone who takes a fresh look at the literature on  the glass transition cannot fail to be struck by the wide diversity of views and approaches. There is a broad consensus on the fact that understanding supercooled liquids, glasses and glass formation represents a deep, interesting, mysterious, and fundamental problem... yet, there is  no general agreement concerning what actually makes the problem deep, interesting, mysterious, and fundamental, nor on the paths to be explored to solve it.  Even worse, there does not seem to exist among the scientific community a shared view of what kind of theoretical achievement would be needed to declare the problem solved. As a result, claims that the problem has already been solved are not uncommon, but encounter a high level of skepticism.

The main purpose of the present review is neither to detail the main features of the phenomenology of glass formation nor to thoroughly assess the validity of the proposed theoretical descriptions, a necessarily very subjective exercise anyhow. Many such reviews have already been published~\protect\shortcite{angell-review95,reviewEAN96,deben-book,ngai,tarjusreview01,deben01,albareview01,cavagnareview09,ludoreview09}\footnote{Topical reviews on a variety of theoretical approaches will be mentioned further down in the text.}  and there is no point in adding one more opus of the sort. I would rather like to give the reader a few indications to get oriented in the apparent jumble formed by the various approaches. Along the way I will also try to identify intrinsic features of the phenomenology which may represent obstacles to a full resolution of the problem and important questions which remain unsettled. 

Before moving on to a more detailed presentation, it is worth stressing that some form of consensus does exist on at least a few points. For instance,  it is by now established that the observed glass ``transition''  is not a \textit{bona fide} phase transition, but rather a dynamical crossover through which a viscous liquid falls out of equilibrium and appears solid on the experimental time scale. (I will nonetheless use the term glass transition without quotation marks in what follows as this is common practice, but the reader should keep in mind the present warning.) The glass transition temperature $T_g$ at which this, indeed quite sharp, crossover occurs depends on cooling rate or observation time;  it is fixed by some operational convention and, typically, corresponds to a viscosity that reaches $10^{13}$ Poise or a relaxation time that is of the order of $10^2$ seconds.  Another widely accepted starting point is that equilibrium statistical mechanics is the relevant framework to describe glass-forming systems, including supercooled liquids which are in a metastable state compared to the crystal; the latter is therefore excluded from theoretical descriptions of glass formation. (This seems reasonable but  may sometimes require some caution: see \protect\shortcite{cavagnareview09} for a pedagogical discussion.)

Considered with some historical perspective, the field seems to be only very slowly evolving. Most of the theories and concepts that are at the forefront of nowadays discussions about the glass transition were around by the mid-eighties. Besides the defect-diffusion \protect\shortcite{glarum}, free-volume \protect\shortcite{freevolume60_a,freevolume60_b} and configurational-entropy \protect\shortcite{gibbs58,gibbs60,adam65} approaches that date back to the sixties, this is for instance the case for the energy-landscape picture \protect\shortcite{stillinger83}, the mode-coupling \protect\shortcite{bengtzelius84,leutheusser84} and the random first-order transition \protect\shortcite{wolynes85,kirkpatrick-wolynes87_a,kirkpatrick-wolynes87_b} theories, the concepts of kinetic constraints \protect\shortcite{palmer84,FA84} and of geometric frustration \protect\shortcite{kleman79,sethna83,nelson83_a,nelson83_b,nelson_etal_a,nelson_etal_b,sadoc84}, all  approaches that will be discussed in more detail below. Similarly, many of the experimental advances were operational twenty years ago. This observation certainly generates a distressing feeling that the glass transition problem, one of the oldest puzzles in physics, does not get any closer to a resolution. However, there is room for optimism. First, new ideas and standpoints do emerge. This is illustrated by the relatively recent surge of experimental, numerical and theoretical work on a previously overlooked aspect of the relaxation in glass-forming systems: the increasing heterogeneous character of the dynamics and the associated growth of space-time correlations as one approaches the glass transition, a property that represents the central topic of this book. In addition, progress has been made along several lines of research, with the development of models and theoretical tools, and sometimes of crisper and possibly testable predictions.

\section{A diversity of views and approaches}
\label{sec:2}

\subsection{What is there to be explained?}
\label{subsec:2.1}
What is there to be explained about the glass transition? What is the scope of the description, \textit{i.e.} what is the range of phenomena and of systems to be included? Answering those questions is already a perilous task that requires  \textit{a priori} choices. The word ``glass'' and the qualifier ``glassy'' are used in a great variety of contexts to describe systems with unusually sluggish dynamics, in which the degrees of freedom of interest remain (apparently) disordered. Such systems can get stuck in an arrested state when the dynamics becomes too slow to be detectable. Some generic phenomena are then observed in a slew of apparently unrelated materials. As an illustration, consider  what goes under the term of  ``aging'' \protect\shortcite{struik,aging98}. Aging describes the fact that the properties of a system depends on its  ``age'', \textit{i.e.} on the elapsed time since it has been prepared. This is most clearly seen in two-time response and correlation functions which, in the aging regime, lose the time-translation invariance found in equilibrium states and depend on the waiting time between preparation of the system and beginning of the measurement. This aging phenomenon, in which an apparent relaxation time is set by the waiting time, is encountered in very different materials; to list a few: polymer glasses (plastics), molecular glasses, colloidal gels, foams, spin glasses, vortex glasses, electron glasses. Aging therefore appears as a ``universal'' property of all out-of-equilibrium systems whose relevant degrees of freedom look frozen on the experimental time scale while still exhibiting some residual motion.  In the above examples, the ``glassy'' degrees of freedom are  associated with molecules, monomers, colloidal particles, bubbles, spin magnetic dipoles, vortices or electrons, and  ``glassiness''  can either be self-generated or result  from the presence of quenched disorder due to impurities and defects. At this level of generality,  it is unclear whether one should look for a basic principle of out-of-equilibrium dynamics or first subdivide the systems into classes characterized, say, by different aging exponents.

While it can certainly be a worthy strategy to try solving the ``glassy" problem in its full blown but rather ill-defined generality, I will take here a more restrictive view that focuses on the glass transition of liquids and polymers. In any case, this can be taken as a starting point. Other systems and more phenomena could be fruitfully brought into the picture, but, in my opinion,  only to the extent that their behavior is argued to closely resemble that of glass-forming liquids and polymers. Needless to say that this approach may not be shared by the whole ``glass community''.

Viewed from the ``liquid side'', \textit{i.e.} upon cooling a liquid (or, possibly, compressing it, see below), the phenomenon of glass formation stands out by the spectacular slowing down of relaxation and the related increase of viscosity that both take place in a continuous manner as temperature is decreased: one observes changes of up to $14$ orders of magnitude in the main ($\alpha$) relaxation time and the viscosity for a $30\%$ variation in temperature. In the same temperature range, the average static correlations between pairs of molecules in the liquid, as probed through static structure factors, barely change. Provided of course crystallization is bypassed, this rather unique feature, which combines a dramatic evolution of the dynamics and an apparently very modest structural change, is seen in virtually all kinds of liquids (inorganic and organic, ionic salts and metallic alloys, polymer melts), liquids with a variety of molecular shapes and a wide range of intermolecular forces.

The phenomenology of glass-forming liquids and polymers is characterized by salient generic trends as temperature decreases. Besides the already mentioned dramatic increase of $\alpha$ relaxation time and viscosity, whose temperature dependence is typically described by a faster-than-Arrhenius form, and the concomitant bland behavior of the static structure factors, the main qualitative features can be summarized as follows: a marked nonexponential time dependence of the relaxation functions (or, equivalently, a marked non-Debye behavior of the frequency-dependent susceptibilitites), the appearance and the development of several relaxation regimes, an increasingly heterogeneous character of the dynamics, with the coexistence over an extended period of time of fast and slow regions, a significant decrease of the entropy, with the difference between the entropy of the supercooled liquid and that of the corresponding crystal dropping by up to a factor of $3$ between the melting and the glass-transition temperatures. (Again, because  the way to present the phenomenology always comes with a preconceived picture, some researchers would probably want to add a few items to the list of central features.)

The above trends and dependences on temperature can be fitted, with reasonable accuracy, by means of various  functional forms. This fact supports the generic and ``universal'' character of the glass-forming behavior. The fitting formulas on the other hand include temperature-independent but material-dependent parameters that contain the specific and nonuniversal aspects of the problem. For instance, the degree of superArrhenius behavior, \textit{i.e.} of departure from Arrhenius temperature dependence of the $\alpha$ relaxation time or the viscosity, is material specific: it is embodied in several (alternative) indices that quantify how  ``fragile''\protect\shortcite{angell_frag85}  a glass-former is. One may of course dream of being ultimately able to predict these material-specific quantities from a first-principle, microscopic theory. In the mean time and more modestly, it is at least possible to look for empirical correlations among them. Numerous attempts of this kind have been made over the years, with varying degrees of robustness. They can be roughly classified into three groups: correlations among characteristics of the slow dynamics, correlations between slow dynamics and thermodynamics, and correlations between properties of the slow and of the fast dynamics. In the first group, one finds for instance correlations between the fragility (see above and \protect\shortcite{angell_frag85}) and the degree of nonexponential behavior of the relaxation functions (the "stretching" parameter) \protect\shortcite{bohmer93} or between the stretching of the relaxation and the amount of decoupling in the temperature dependences of the translational and the rotational diffusion motions \protect\shortcite{ediger_review00}.  In the second group are the correlation beween the heat capacity jump at the glass transition $T_g$ and the fragility \protect\shortcite{angell_frag85} or else beween the decrease of  the "configurational" entropy (excess entropy of the supercooled liquid over that of the crystal)  and the increase of the relaxation time \protect\shortcite{richert98,correl_angell}. Finally in the third group, one can list many proposed correlations between the fragility and various characteristics of the fast dynamics either in the liquid or the glass (relative amplitude of the Boson peak \protect\shortcite{sokolov93}, Poisson ratio of elastic moduli \protect\shortcite{sokolov04},  nonergodicity parameter \protect\shortcite{scopigno03}, etc).   Whether one discards as fortuitous and wobbly, or else takes seriously some of these correlations is strongly connected to the type of description of the glass transition one is aiming at.

\subsection{Sorting out the theoretical approaches}
\label{subsec:2.2}

Even with the restriction that one focuses on what governs glass formation in liquids and polymers, there is plenty of room for diversity and controversy. A central question, to which different answers are given and which is therefore far from purely rhetorical, is whether a general theory of glass formation is possible? By ``general theory''  is meant here a theory that explains the main features associated with the glass transition, without having to rely on a full description of the microscopic details that vary from one substance to another \protect\shortcite{kiv-tar08}. A corollary to the existence of such a theory is that the phenomenon displays some form of  ``universality''. Although challengeable, a number of elements point toward a positive answer, which most certainly explains the attraction that the field has exerted on theorists over the years.

As briefly summarized above, many pieces of the phenomenology of glass-forming liquids appear general enough to be described by master curves (with admittedly a number of material-specific adjustable parameters) and empirical correlations are (more or less successfully) established beween various properties. All of this suggests some common underlying explanation. Beyond this, the dramatic superArrhenius rise of the $\alpha$ relaxation time with decreasing temperature, which for a given glass-former is very similar in a wide variety of dynamical measurements (with a few well identified and interesting exceptions corresponding to a decoupling of the motions either of different entities, as the decoupling between conductivity and viscosity in ionic mixtures \protect\shortcite{reviewEAN96}, or of the same entities probed on very different length scales, as the decoupling between local rotational relaxation and translational diffusion in molecular liquids and polymers \protect\shortcite{ediger_review00,sillescu99}), seems indicative of growing collective behavior. Such a growth is of course what partly washes out the influence of molecular details, which can then be incorporated in effective parameters, and can lend itself to a general theory.

Attributing the viscous slowing down in supercooled liquids and polymers to a collective phenomenon is at the core of most theories. Collective behavior implies the existence of at least one supermolecular length scale that can be associated with growing correlations or growing coherence in the system as temperature decreases. This issue of length scales has indeed become central in most recent discussions and developments concerning the glass transition. More will be said below on this, but for now it is worth recalling a few points. First, the nature of the putative supermolecular lengths  characterizing the slowdown of relaxation and flow is quite elusive. Nothing much happens in the static pair correlations, which are the only measures of the structure that are easily accessible experimentally. As a consequence, if spatial correlations of one sort or another grow, they must be of rather subtle and unusual nature. The experimental observation that has recently put length scales at the forefront is that of the heterogeneous character of the dynamics. It has been shown through both numerical and experimental work that the dynamics becomes spatially correlated as one approches the glass transition, with an emerging length scale related to the typical size of the dynamical heterogeneities (see \protect\shortcite{ediger_review00,sillescu99} and the other chapters of this book). It is not clear however if this  ``dynamical'' length is the one controlling the viscous slowing down. Another keypoint, and a most annoying fact, is that, irrespective of their precise definition, supermolecular lengths in glass-forming systems never seem to grow bigger than a few nanometers,  \textit{i.e.} $5$ to $10$ molecular diameters at most \protect\shortcite{ediger_review00,chi_T_a,chi_T_b,richert10}. Collective behavior is thus present, but it may never fully dominate local, molecular effects in the experimentally accessible range. The absence of a singularity with diverging time and length scales that could be seen or closely approached in experiments can be viewed as an intrinsic difficulty of the glass transition problem.

Looking for a general theory does not exhaust the possible lines of research on glass formation. For instance, one may want to know how the processes by which molecules move change between the high-temperature liquid and the glass transition  or to describe the viscous slowing down in terms of specific mechanisms for relaxation and flow. Universality, considered in a weaker sense than when associated with diverging length scales at critical points, could still be observed if a well identified mechanism prevails. This would then be more akin to the dislocation description of plastic motion for crystals \protect\shortcite{dislocation_a,dislocation_b}. Finally, it could also well be, but this would be disappointing to many, that understanding glass formation lies in chemistry with no way out of microscopic and substance-specific computations or measurements, the generic character of the phenomenology summarized above being then only superficial.

To  contrast the various theoretical approaches of the glass transition, it is helpful to sort out their starting points and perspectives according to  pairs of related opposites:

(i) \textit{Liquid side versus amorphous-solid side.}

Above, I have emphasized the glass transition when approached from the liquid side, as it shows the most remarkable pieces of phenomenology. However, viewed from the  ``solid side'', glass formation may be quite fascinating too. What makes the rigidity of an amorphous phase devoid of long-range order?, what is the specificity of the associated vibrational modes and elastic response?, how does the glass  ``melt''?  are all challenging questions (see \textit{e.g.} \protect\shortcite{wyart05}). The glass is for sure a frozen liquid, but it has also been suggested to envisage the viscous liquid as a  ``solid that flows'' \protect\shortcite{dyre06}.

(ii) \textit{Coarse-graining, scaling and underlying critical point versus atomic-motion level and relaxation mechanisms.}

Growing collective behavior is a well established cause of emerging universality, and it is then tempting to apply the recipes that have been very successfull in statistical physics: coarse-graining procedures and use of effective theories, search for critical points around which to organize scaling analysis. Being biased toward the possibility of such a general theory of the collective behavior associated with glass formation, I may somewhat overlook in the following the numerous ``microscopic'' approaches that focus on describing the actual motion of the molecules and the way it changes as temperature is decreased. Such detailed information is hardly obtainable from experiments, with the possible exception of colloidal suspensions and driven granular media, but is provided by computer simulations \protect\shortcite{walter}. In addition, as alluded to above, it may also well be that the apparent universality of the slowdown of dynamics leading to glass formation originates from the predominance of a specific relaxation mechanism rather than from an underlying critical point.

(iii) \textit{Real three-dimensional space versus configurational space.}

That relaxation in a viscous liquid seems well described in terms of thermally activated events suggests that the system may be temporarily trapped in  ``metastable states'' and, accordingly, that a fruitful framework is provided by a description of the  ``energy landscape'' and its topographic properties \protect\shortcite{goldstein69,stillinger83,deben01}. This configurational-space picture, which focuses on the hypersurface formed by the potential energy or some coarse-grained free-energy as a function of all configurational variables, is one of the paths followed to describe glassy systems. Alternatively, one may attempt a real-space description of the phenomena. For instance, activated processes correspond to rare, localized events that are more easily describable from a real-space than from a configurational-space perspective. A similar conclusion applies to the dynamical heterogeneities.
 
(iv) \textit{Kinetics versus statics.}

The observed glass transition is undoubtedly a kinetic crossover and the most spectacular phenomena, slowing down and dynamical heterogeneities, pertain to the dynamics. Yet, on general ground (see below), one may expect some thermodynamic and structural underpinning.  Whether the physics of glass formation can be understood on the basis of a purely dynamical origin with no thermodynamic signature or requires a thermodynamic or structural explanation represents a central issue. The case for the latter option is  \textit{a priori} difficult to pin down due to the absence of clear evidence showing growing thermodynamic or structural correlations as one approaches the glass transition. (It can of course be argued that such correlations are subtle, and consequently hard to detect, and that their growth is limited, with an impact on the dynamics that is enormously amplified.) On the other hand,  a purely kinetic explanation overlooks the thermodynamic aspects of the phenomenology, such as the rapid decrease of the ``configurational'' entropy.
\\

Before closing this section, I would like to come back to the theories that relate the apparently universal features of glass formation to some collective behavior controlled by underlying phase transitions. As no such transitions are observed in experiments, the putative singularities must be located outside of the physically accessible range of parameters: the postulated critical points and phase transitions may then be either \textit{avoided} or  \textit{unreachable}. ``Avoided'' implies that the singularity only appears as a crossover in real life, at a temperature that is above the experimental $T_g$;  ``unreachable'' means that the transition occurs at a temperature below $T_g$, being for that reason inaccessible in experiments. In line with point (iv) made above, the nature of the hypothesized critical points and transitions can moreover be either dynamic or static. A more detailed description along these elements of classification will be provided in sections \ref{sec:4} and \ref{sec:5}.

\section{Elements of theoretical strategies}
\label{sec:3}

As briefly sketched in the preceding sections, the pursuit of a theoretical description of the glass transition is multiform. An overview of theories should then try to do justice to the diversity of  concepts, models, tools, ideas, and frameworks that have been put forward in the context of glass-forming systems. The motivation for such an exercise is not only oecumenical. First, at the present time, no theoretical approach has been accepted as the definite answer to the glass problem: different paths may ultimately prove fruitful and complementary. Second, some of the constructs devised in the course of studies on glass-formers may find a life of their own and be transposable to other fields. The list given below does not pretend to be exhaustive.

\subsection{Minimal models and simplifying concepts}
\label{subsec:3.1}

An obvious strategy in physics to address a seemingly general phenomenon is trying to identify the main ingredients and get rid of all superfluous microscopic details. Through some level of coarse-graining, choice of relevant degrees of freedom and use of simplifying concepts, minimal models and effective theories can then be proposed as starting points for further studies. To pick up a well known example in a closely related field: the study of spin glasses has been put on a firm basis with the proposal by Edwards and Anderson of their lattice hamiltonian model. In this case, two physical ingredients, frustration and quenched disorder, have been incorporated in a simple classical Ising hamiltonian \protect\shortcite{EA75}. The order parameter and the associated phase transition are then adequately described (at least at a minimum level), and the detailed nature of the spin degrees of freedom (except for symmetry), of the interactions among the latter, of the distribution of the quenched disorder, and of the underlying lattice is essentially irrelevant to an understanding of the spin-glass phenomenology. This, by far, does not mean that the spin-glass problem is solved. Central questions concerning the Edwards-Anderson model, such as the nature of the spin-glass phase and the existence of a transition in nonzero magnetic field, are still awaiting a definite answer. But the starting point is well accepted.

The situation is unfortunately not as clear for glass-forming liquids.  There are models of liquids, but they can hardly be considered as minimal models for a theory of glass formation. This is for instance the case of the Lennard-Jones models for atomic liquid mixtures and of the hard-sphere models for repulsive colloidal suspensions, which are at the heart of most computer simulation studies \protect\shortcite{walter}. Such ``realistic'' atomistic models still incorporate a wealth of microscopic details without allowing one to tune in any significant way the salient features of glass formation such as the fragility.

To build minimal models or effective theories of glass-forming systems, several routes have been proposed. They rely on various ingredients and concepts that are hypothesized to account for the origin of the increased sluggishness and of the heterogeneous character of the relaxation as temperature decreases, be they cooperativity, free volume, jamming, kinetic constraints, facilitation, local packing constraints, geometric frustration, etc. They may also involve analogies with spin-glass and other disordered models \protect\shortcite{kirkpatrick-wolynes87_a,kirkpatrick-wolynes87_b,kirkpatrick-thirumalai_a,kirkpatrick-thirumalai_b,sethna91,moore02}, uniformly frustrated models \protect\shortcite{nelson02,frustratedXY,grousson_a,grousson_b}, etc. Some of these attempts will be further discussed below.

\subsection{Looking for a localized relaxation mechanism in real space}
\label{subsec:3.2}

That a full understanding of glass formation will ultimately  require knowledge of the principal relaxation mechanism(s) is rather undisputed, as is the fact  that relaxation in glass-forming liquids proceeds via localized events in space.  On the other hand, looking for such mechanisms may not necessarily be considered as the highest priority for building a theory. Here, however, I more specifically refer to those approaches that put emphasis on mechanisms without invoking underlying phase transitions.

A line of research indeed posits that understanding glass formation only requires identifying a dominant relaxation mechanism that is supposed to take over as the liquid becomes more viscous. Examples taking this perspective include the excitation-chain \protect\shortcite{langer_a,langer_b} and single-particle barrier-hopping \protect\shortcite{schweizer}  approaches, the description in terms of  quasi-particles \protect\shortcite{procaccia09}, and the elastic models \protect\shortcite{dyre06}. In the ``shoving model'' \protect\shortcite{dyre06} for instance, localized flow events are assumed to take place in compact regions whose rearrangement involves an increase of volume. The activation energy then comes from the elastic cost associated with  ``shoving''  the rest of the system, which is taken as a solid on the short time scale of the event. In this work, the superArrhenius increase of the activation energy is related to the temperature dependence of the infinite-frequency shear modulus. Slowing down occurs with no growing length scale (hence with no underlying singularity), since the typical size of the rearranging regions stays constant  with decreasing temperature.

It is also worth noting that computer simulations of atomistic models offer an opportunity to look for the main mechanisms for relaxation and flow in real space. In this vein, most simulation studies on the heterogeneity of the dynamics in glass-forming liquids have focused on dynamical clustering, providing some evidence for strings and micro-strings \protect\shortcite{string98,microstring04}, or other types of clusters \protect\shortcite{democratic06}. More recently, Harrowell and coworkers \protect\shortcite{harrowell_a,harrowell_b} have investigated the collective moves of the atoms, in the form of strain events, localized reorganizations, soft modes, etc, that could be responsible for the irreversible nature of structural relaxation and flow in supercooled liquids.  As already discussed, these attempts could provide a description of glass-forming liquids at a level which is comparable to that of plastic flow in crystalline materials by the dislocation-based theory.

\subsection{Developing a specific statistical-mechanical framework}
\label{subsec:3.3}

As stressed in the Introduction, there is a general agreement concerning the fact that supercooled liquids, and more generally glass-forming systems above their glass transition, can be treated by equilibrium statistical mechanics (provided that one excludes the crystal phase). In this general setting, specific frameworks have been devised to study glass-formers. Interestingly, these theoretical constructs developed in the context of glasses may prove useful, or even have already proven so,  in other scientific fields. I sketch some of these developments below.

One such framework rests on a ``topographic'' view of the phenomena associated with glass formation. The core object is the ``landscape'' formed by the potential energy surface plotted as a function of the coordinates of all the atoms in the liquid. This leads to a reformulation of statistical mechanics which has been put forward by Stillinger and coworkers \protect\shortcite{stillinger83,stillinger95,deben01}. Focus is shifted to a study of the properties of the (temperature-independent) landscape, \textit{i.e.} the statistics of the minima and the saddle points, the characteristics of their neighborhoods, the connectivity graph between minima, etc \protect\shortcite{wales}. The potential descriptive power of the approach comes from the remark, nicely articulated in \protect\shortcite{goldstein69}, that the viscous liquid at low enough temperature spends most of the time vibrating around typical energy minima (later called ``inherent structures''  by Stillinger \protect\shortcite{stillinger83}) with only rare, localized reorganization events that are associated with thermally activated barrier hopping between minima. The potential-energy landscape formalism then provides a vista for rationalizing, in a qualitative way, collective phenomena occuring in supercooled liquids \protect\shortcite{stillinger95}. It also represents a framework in which to compute properties of specific systems by means of numerical simulations \protect\shortcite{keyes92,sastry98,sciortino05,heuer08}. The daunting difficulty of course comes from the dimensionality of configurational space. With its $3 N$ dimensions, $N$ being the number of atoms, it makes the landscape technically hard to characterize and, in effect, conceptually hard to visualize.

With some (usually unspecified) coarse-graining in mind, it may be fruitful to go from an energy landcape to a free-energy one, in which energy minima separated by small barriers are grouped into a free-energy state. Such a construction is well-defined at a mean-field level, and classes of complex free-energy landscapes with multiple metastable states have been found and thoroughly characterized in theoretical studies of systems with quenched disorder, mostly mean-field spin-glass models\protect\shortcite{pedestrianSG}. Supercooled liquids and the associated glasses, however, have no quenched disorder. Remarkably, some powerful tools introduced in the context of spin glasses\protect\shortcite{disordered}, such the replica method and generating functional approaches, have been generalized to investigate, at a mean-field level at least, the complex free-energy landscape of glass-forming liquids\protect\shortcite{monasson95,pedestrianSG,review-replicas,functional}. This point will be further discussed in section \ref{subsec:5.2}.

The two other formalisms that I would like to briefly mention also represent, if not reformulations, at least new twists in statistical mechanics. The first one is the ``iso-configurational ensemble'' \protect\shortcite{widmer_a,widmer_b} which is formed by all trajectories that start from an identical configuration of particles with random initial velocities sampled from an equilibrium Maxwell distribution. Together with such notions as ``dynamic propensity'', it has been introduced to study whether the static structure of a glass-forming liquid, without a priori knowledge of which of its features might be relevant,  influences the heterogeneous character of the dynamics \protect\shortcite{widmer_a,widmer_b}. The second formalism goes somewhat beyond the realm of equilibrium statistical mechanics:  it considers the space of trajectories and introduces an (\textit{a priori} unphysical) external field that couples to the mobility of the system in such a way that it can drive the latter out of equilibrium in an immobile, nonergodic phase \protect\shortcite{nonequil_a,nonequil_b}. The presence of a nonequilibrium transition between ergodic and nonergodic phases can then be investigated either in idealized models or in atomistic ones (see also section \ref{subsec:4.2}).

\subsection{In search of a growing length scale}
\label{subsec:3.4}

For those theories that associate glass formation with some sort of universality and emerging collective behavior, a keypoint is the characterization of a length which grows as temperature is decreased and which is directly connected to the slowdown of relaxation. A related, but not quite coincident, issue is that of the existence of a length associated with the increasing spatial heterogeneity of the dynamics.

Recently, there has been a major effort to go beyond heuristic or \textit{ad hoc} definitions of  length scales in glass-forming liquids, such as the cooperativity length introduced in the Adam-Gibbs postulate \protect\shortcite{adam65} of cooperatively rearranging regions (see below) or various measures of dynamical clustering. This has led to introducing appropriate  correlation functions which are in principle computable.

Much progress has been made concerning the growing spatial correlations in the dynamics. It has been realized that the latter, which are associated with fluctuations around the averaged dynamics and with the phenomenon of dynamical heterogeneities, are describable through multi-point space-time correlation functions. Information on the corresponding ``dynamical'' correlation length can be extracted from a 4-point correlation function that describes how far the dynamics at a given point in space affects the dynamics at another point and, with some plausible assumptions, from the associated dynamic susceptibility \protect\shortcite{dasgupta91,silvio-glotzer_a,silvio-glotzer_b,chi_JCP_a,chi_JCP_b}. Direct or indirect evidence for a ``dynamical'' length that grows as temperature decreases has been obtained in this general framework. The topic being at the heart of the present book and developed in the other chapters, I will not dwell more on it, except to mention an interesting result obtained in this context: it has been shown that the dynamical singularity predicted by the mode-coupling theory of glass-forming liquids comes with a divergence of the above described ``dynamical'' length \protect\shortcite{franz-length,BB-length}, thereby quashing a previously widespread belief that the relaxation time diverges with no accompanying diverging length scale.

A separate issue, which has deep consequences on the theoretical picture of glass-forming systems, is whether there also exists a growing static length as temperature decreases. If there is a true divergence of the $\alpha$ relaxation time at a nonzero temperature (a hypothesis that of course cannot be verified experimentally), heuristic and rigorous arguments \protect\shortcite{montanari06,semerjian} prove that it must come with the divergence of a static correlation length, albeit a complicated one involving point-to-set correlations. If, on the other hand, the relaxation time only diverges at zero temperature, the situation is not as clear-cut. One could envisage a relaxation process whose temperature behavior is characterized by an Arrhenius behavior with a purely local energy barrier, so that no static correlations grow as the relaxation time increases (and diverges at zero temperature). It would therefore seem more relevant to study the behavior of the relaxation time normalized by the local, ``bare'' relaxation time: $\tau(T) / \tau_{0}(T)$ with \textit{e.g.} $\tau_{0}(T) \sim \exp(\frac{E_{\infty}}{T})$. With the assumption that the rigorous bounds between length and time scales derived by Montanari and Semerjian \protect\shortcite{montanari06} are of general validity, one then concludes that the growth of this scaled relaxation time must come with a growing static correlation length. This suggests that the superArrhenius temperature dependence of the $\alpha$ relaxation time of fragile glass-forming liquids, for which, even after normalization by a ``local'' Arrhenius-like relaxation time, $\tau(T)/\tau_{0}(T)$ behaves as $\exp(\frac{\Delta E(T)}{T})$ with $\Delta E(T)$ increasing  as temperature decreases, is indicative of collective behavior with a concomitantly growing static correlation length.

There are several possible, and not mutually excluding, strategies to search for static correlations associated with glass formation. Since no interesting  correlations  show up in simple $2$-body structural measures such as the static structure factor, one may then think of: (i) studying how far amorphous boundary conditions influence the system, thereby looking at static point-to-set correlations such as those discussed above \protect\shortcite{BBdroplet04,montanari06,silvio_kac07,pt-to-set07,pt-to-set08,cavagnareview09}, (ii) using finite-size analysis for chosen thermodynamic quantities \protect\shortcite{heuer03,finitescaling06,finitescaling09}, (iii) looking at a crossover size in pattern repetition within a configuration \protect\shortcite{kurchan09}, or else (iii), provided that the locally preferred arrangement of the molecules in the liquid has been properly identified, investigating the static pair correlations of the associated local order parameter, which amounts to considering multi-particle correlations as in bond-orientational order parameters \protect\shortcite{steinhardt81,tanaka_a,tanaka_b,sausset10}. Up to the present at least, these procedures are not experimentally realizable in liquids. However, they have been numerically tested on  liquid models with encouraging preliminary results concerning the increase of a static length with decreasing temperature.

\section{Theories based on an underlying dynamical transition}
\label{sec:4}

As discussed above, glass formation takes place in liquids and polymers without any observed singularity. Universality and detail independence, if indeed a genuine property of the glass transition phenomenon, are explainable on the basis of underlying critical points that control the physics of the viscous slowing down but are either avoided or unreachable. In this section, I will discuss theories that involve purely dynamical transitions with no thermodynamic signature.

\subsection{Mode-coupling theory: an avoided dynamical transition at $T_c > T_g$}
\label{subsec:4.1}
	
The \textit{mode-coupling} theory of glass-forming liquids predicts a dynamical arrest without any significant change in the static 
properties \protect\shortcite{gotze_a,gotze_b,gotze-sjogren}. 
The latter are assumed to behave smoothly, and the viscous slowing down results from a nonlinear feedback mechanism affecting the relaxation of the density fluctuations. Formally, the theory involves a set of nonlinear integro-differential equations describing the evolution of the dynamic structure factor $S(q,t)$, which is the wave-vector- and time-dependent pair  correlation function of the density fluctuations. These equations have been originally derived by using the Zwanzig-Mori projection-operator 
formalism \protect\shortcite{gotze_a,gotze_b,gotze-sjogren}, 
but they can be obtained as well within a dynamical 
field-theoretical framework \protect\shortcite{mazenko_a,mazenko_b,miyazaki05,kawasaki08,andreanov09}. 
The crux of the approach consists in formulating an approximation, 
the  ``mode-coupling approximation'',  that allows one to close formally exact dynamical equations and write down a tractable 
set of self-consistent equations for $S(q,t)$. The key input that comes into the approximate theory is the static structure factor, $S(q)\equiv S(q,t=0)$.

The solution of the self-consistent equations predicts a slowdown of the relaxation of $S(q,t)$ with decreasing temperature that is physically attributed to a  ``cage effect'' and to the feedback mechanism above mentioned. This solution exhibits a dynamical freezing at a critical point $T_c$ which represents a transition from an ergodic to a nonergodic state with no concomitant singularity in the thermodynamics of the system \protect\shortcite{bengtzelius84,leutheusser84,gotze_a,gotze_b}. The $\alpha$ relaxation time diverges in a power-law fashion for $T>T_c$,
\begin{equation}
\tau(T) \sim (T-T_c)^{- \gamma},
\end{equation}
and several specific predictions are made concerning the scaling behavior near to $T_c$.

Early on,  it was realized that the dynamical arrest at $T_c$ could not describe the observed glass transition at $T_g$, nor a transition to a putative ideal glass at a temperature below $T_g$, and that $T_c$ should rather be located above $T_g$. The singularity at $T_c$ must then be interpreted as ``avoided'' and manifesting itself as a crossover in the phenomenolgy of glass-forming liquids \protect\shortcite{gotze_a,gotze_b}. The mode-coupling approach can thus at best describe the dynamics of moderately supercooled liquids, for which its main achievement is the predicted appearance of a two-step relaxation process as temperature decreases, as indeed observed in experiments and in simulations. The mode-coupling theory has also proven a versatile scheme to study additional systems, such as repulsive and attractive colloids \protect\shortcite{zaccarelli}, and phenomena. It has been for instance generalized  to investigate aging dynamics in the nonergodic phase \protect\shortcite{cugliandolo,aging98}, nonlinear rheology of a variety of glassy systems \protect\shortcite{fuchs_a,fuchs_b}, or more recently dynamical heterogeneities and multi-point space-time correlations \protect\shortcite{chi_JCP_a,chi_JCP_b,IMCT,tarzia10}, all cases for which it provides nontrivial predictions.

What is the nature of the mode-coupling approximation and why is the singularity avoided ? The traditional answers \protect\shortcite{gotze_a,gotze_b,gotze-sjogren} invoking freezing due to a ``local cage effect'' (see above) and avoidance due to ``hopping mechanisms'' are not satisfactory. A clearer picture has emerged from work on \textit{a priori} unrelated systems, mean-field ``generalized'' spin glasses \protect\shortcite{kirkpatrick-thirumalai_a,kirkpatrick-thirumalai_b,cavagnareview09,BB-RFOT09}, and further developments. The analogy with mean-field spin glasses will be discussed in more detail in the section on the random first-order transition theory. To make a long and elaborate story short, let me summarize the findings as follows:  

(i) The mode-coupling approximation and the associated self-consistent equations have a mean-field character \protect\shortcite{KTW89,andreanov-landau09,BB-RFOT09}.

(ii) In a free-energy landscape picture, the ergodicity breaking transition corresponds to the disappearance of unstable directions for escape and to the emergence of infinite barriers between the dominant states \protect\shortcite{cavagnareview09}.

(iii) The divergence of the relaxation time at $T_c$ is not due to a purely local effect, but involves the divergence of a length characterizing spatial correlations in the dynamics \protect\shortcite{franz-length,BB-length}.

(iv) Being mean-field in character, the mode-coupling singularity is affected in real, finite-dimensional, systems by fluctuations whose influence is two-fold \protect\shortcite{andreanov-landau09}. First, there is a standard effect that can often be handled through perturbative expansions:  the exponents describing the scaling behavior are modified below an upper critical dimension, which in the present case is found equal to $d=8$ \protect\shortcite{BB-d_upper,chi_JCP_a,chi_JCP_b,parisi_franz10}. In addition, and with more severe consequences,  rare localized events corresponding to thermally activated processes destroy the singularity; these intrinsically nonperturbative phenomena operate in all finite dimensions.

This puts the mode-coupling approach on a much firmer basis.  The down side is that a major theoretical breakthrough is needed to incorporate nonperturbative effects beyond the mean-field picture (see section \ref{subsec:5.2} below).

\subsection{Dynamical facilitation and kinetically constrained models: unreachable dynamical critical point at $T=0$ and avoided dynamical 
 first-order transition}
\label{subsec:4.2}

As for the mode-coupling theory, the radical perspective taken by the \textit{dynamical facilitation} approach \protect\shortcite{review-facilitation10} is that the main characteristics of glass-forming liquids can be described by purely dynamical arguments. Building on the observation that the static pair correlations change only weakly with temperature and that the associated correlation length always stays of the order of the molecular diameter, this approach indeed assumes that, after some suitable coarse-graining, static correlations become negligible and that, accordingly, thermodynamics is trivial. In this picture, glassiness, cooperativity and heterogeneity in the dynamics result from effective kinetic constraints that emerge at low enough temperature when mobility in a supercooled liquid is concentrated in rare localized regions, the rest of the system being essentially frozen. Such ``mobility defects'' are taken as the effective degrees of freedom which, together with the postulated kinetic rules that constrain their motion, form the basis of the theoretical description. ``Facilitation'' in this context means that mobility defects trigger mobility in neighboring regions. 

This general scenario has led to the formulation of  families of models generically referred to as ``kinetically constrained models''\protect\shortcite{KCMreview}. These models rest on a hamiltonian for noninteracting variables (spins or particles on a lattice) combined with specific constraints on the allowed moves of any such variable. These constraints involve a dependence on the local neighborhood, \textit{e. g.} a particle can hop to a different site only if the number of nearest neighbors at the original and the target sites is less than a fixed threshold value. These models are sufficiently tractable for allowing detailed analytical and numerical calculations. It has been shown that cooperativity of the relaxation, with a superArrhenius temperature dependence of the relaxation time, is a property of many of the models and that heterogeneity of the dynamics naturally emerge as a central feature of all models. The kinetically constrained models qualitatively reproduce many aspects of the slow dynamics of glass-forming liquids, but, more specifically, they provide a consistent picture of the dynamical heterogeneitites in a space-time setting \protect\shortcite{gar-chand02,KCMreview,review-facilitation10}.

Beyond the variety of models and associated behaviors, it has been suggested that the apparent universality of the dynamical properties in glass-forming systems could come from the existence of a dynamical critical point at zero temperature \protect\shortcite{RG_Whitelam04_a,RG_Whitelam04_b,RG_KCM,nonequil_a,nonequil_b}. Dynamical scaling analysis can then be organized about this zero-temperature singularity. For instance, inspired by the low-$T$ behavior of some ``hierarchical'' kinetically constrained models\protect\shortcite{hierarch91}, Garrahan and Chandler \protect\shortcite{gar-chand03,review-facilitation10} have proposed to describe the superArrhenius temperature dependence of glass-forming liquids by a B\"{a}ssler-type expression \protect\shortcite{bassler87}, 
\begin{equation}\label{eq:34}
\tau(T) \sim \exp \left( \frac{A}{T^2} \right),
\end{equation}
for temperatures much below an ``onset'' that marks the begining of facilitated dynamics with rare mobile regions.

The facilitation approach puts the emphasis on a space-time picture of glassy dynamics. At a descriptive level, dynamical heterogeneities in glass-forming liquids can be for instance associated to the proximity to a first-order nonequilibrium transition in trajectory space which is characterized by the coexistence of a mobile and an immobile phase \protect\shortcite{nonequil_a,nonequil_b} (see section \ref{subsec:3.4}).

As most theories (with the possible exception of the mode-coupling theory discussed above), the present approach comes up against the difficulty of deriving rather idealized models from realistic systems. This is by no means benign. For instance, the simple assumption that mobility is conserved,\textit{ i.e.} that there is no spontaneous appearance or disappearance of mobility defects, is questionable \protect\shortcite{candelier09}. More importantly, the facilitation approach cannot address the aspects of the glass-forming phenomenology that involve thermodynamics (\textit{e.g.}, the behavior of the entropy and of the heat capacity \protect\shortcite{defectsBBT05}) nor the nontrivial static correlations that are argued to accompany the increase of relaxation time in fragile glassformers (see above).

\section{Theories based on an underlying thermodynamic or static transition}
\label{sec:5}

In this section will be considered theoretical approaches that relate the (hypothesized) collective behavior of glass-forming systems and associated universality to underlying thermodynamic or static critical points. I will stress those theories that are amenable to a genuine statistical-mechanical treatment and involve effective hamiltonian models. However, before doing so, it is worth mentioning two phenomenological pictures, the free-volume and the configurational-entropy models, as they have been influential for the thinking on the glass transition, and are still commonly used to rationalize data on liquids and polymers at a semiempirical level.

\subsection{Free-volume and configurational-entropy models:  unreachable transition points at $T_0 < T_g$}
\label{subsec:5.1}

\textit{Free-volume} models rest on the assumption that molecular transport in viscous fluids occurs only when voids having a volume large enough to accommodate a molecule form by the redistribution of some ``free volume'' \protect\shortcite{freevolume60_a,freevolume60_b,freevolume}. The latter is loosely defined as some surplus volume that is not taken up by the molecules. In the standard presentation, a molecule in a dense fluid is mostly confined to a cage formed by its nearest neighbors and the local free volume $v_f$ is that part of a cage space which exceeds the volume taken by a molecule. It is then assumed that between two events contributing to molecular transport, a reshuffling of free volume among the cages occurs at no cost of energy and that the local free volumes are statistically uncorrelated. This leads to an expression for the  viscosity,
\begin{equation}
\label{doolitle}
\eta(T)=\eta_{0}\exp \left( \frac{K}{v_f(T)} \right),
\end{equation}
with $K$ essentially constant, which is also known as the Doolittle equation \protect\shortcite{doolittle}.

The free-volume mechanism fundamentally relies on a hard-sphere picture in which thermal activation plays no role. For application to real liquids and polymers, temperature enters through the fact that molecules and monomers are not truly hard and that the constant-pressure volume is temperature-dependent. An underlying transition comes into play when further assuming that all free volume is consumed at a nonzero temperature $T_0 < T_g$, which, when inserted in Eq.~(\ref{doolitle}), gives the Vogel-Fulcher-Tammann (VFT) expression,
\begin{equation}
\label{VFT}
\eta(T)=\eta_{0}\exp \left( \frac{D T_0}{T-T_0} \right),
\end{equation}
also called Williams-Landel-Ferry (WLF) formula in the context of polymers and widely used to fit experimental data.

The \textit{configurational-entropy} picture on the other hand rests on the idea, popularized by Goldstein \protect\shortcite{goldstein69},  that relaxation in a deeply supercooled liquid approaching the glass transition is best described by invoking motion of the representative state point of the system on the potential energy hypersurface \protect\shortcite{gibbs60} (see section \ref{subsec:3.3}). In this view, the slowing down of relaxation and flow with decreasing temperature is related to a decrease of the number of available minima and of the associated ``configurational entropy''. The Adam-Gibbs approach \protect\shortcite{adam65} represents a phenomenological attempt to make this relation more precise. Structural relaxation is assumed to take place through increasingly cooperative rearrangements of groups of molecules. Any such group, called a cooperatively rearranging region, is assumed to relax independently of the others.  The effective activation energy for relaxation is then equal to the typical energy barrier per molecule, which is taken as independent of temperature, multiplied by the number of molecules that are necessary to form the smallest cooperatively rearranging region. This latter number goes as the inverse of the configurational entropy per molecule $s_c(T)$, which leads to the following expression for the $\alpha$ relaxation time:
\begin{equation}
\label{adam-gibbs}
\tau(T)=\tau_{0}\exp \left( \frac{C}{T s_c(T)} \right),
\end{equation}
with $C$ a constant.

If the configurational entropy vanishes at a nonzero temperature $T_0$, an assumption somewhat analogous to that made in the free-volume model (see above)\footnote{The vanishing of the configurational entropy at a nonzero temperature is however found in the Gibbs-di Marzio approximate Flory-Huggins mean-field treatment of a lattice model of linear polymeric chains\protect\shortcite{gibbs58}.}, the relaxation time diverges at this same nonzero temperature. In particular, if the configurational entropy is identified with the entropy difference between the supercooled liquid and the crystal, the Adam-Gibbs theory correlates the extrapolated divergence of the relaxation time at $T_0$ with the extrapolated vanishing of the excess entropy at the so-called Kauzmann  temperature $T_K$ \protect\shortcite{kauzmann48}.  Equation (\ref{adam-gibbs}) also gives back the VFT/WLF formula, Eq. (\ref{VFT}),  by assuming that the configurational entropy vanishes linearly at $T_0$.

Note that in both the free-volume and the Adam-Gibbs configurational-entropy approaches, the unreachable thermodynamic transition temperature $T_0$ does not appear as a necessary ingredient of the theoretical description, but rather as a convenient input to derive the empirical VFT/WLF expression.

\subsection{Random First-Order Transition theory: an unreachable thermodynamic critical point at $T_K < T_g$ coupled with an avoided dynamical transition at $T_c > T_g$}
\label{subsec:5.2}

The \textit{random first-order transition} theory \protect\shortcite{KTW89,wolynesreview07} can be seen as a three-stage construction. The foundations are formed by an intricate mean-field theory of glass formation. In the eighties, Wolynes and coworkers \protect\shortcite{kirkpatrick-wolynes87_a,kirkpatrick-wolynes87_b,kirkpatrick-thirumalai_a,kirkpatrick-thirumalai_b,KTW89} realized that many (postulated) elements of the  description of glass-forming liquids, the Kauzmann-like thermodynamic ``catastrophe'' associated with the extrapolated vanishing of the configurational entropy \protect\shortcite{kauzmann48}, the mode-coupling dynamical singularity and the emergence of a complex (free) energy landscape with a multitude of trapping minima, could all be tied together in a coherent scenario for which explicit realizations were provided by mean-field ``generalized'' spin glasses (such as the random $p$-spin and Potts glass models with infinite-range interactions).

It has been by now established that the scenario is not only realized is spin models with quenched disorder. It is more generally characteristic of a whole class of glass-forming systems described at a mean-field level, with a transition to an ideal glass phase which is second-order in the usual thermodynamic sense (with, \textit{e. g.}, no latent heat) but is accompanied by a discontinuous jump in the order parameter \protect\shortcite{wolynesreview07,cavagnareview09,BB-RFOT09,review-replicas}. This ``random first-order transition'' (also called ``one-step replica symmetry breaking'') phenomenology, with a high-temperature ergodicity breaking transition at $T_c$ and a low-temperature thermodynamic glass transition at $T_K$ that are separated by a regime in which an exponentially large (in system size) number of metastable free-energy states dominates the thermodynamics while trapping the dynamics, has been found in several standard liquid models when treated within mean-field-like approximations \protect\shortcite{wolynes85,mezard97,silvio_HNC,zamponi05,DFT08}. Specific methods, involving generating functionals and replica formalism (see also section \ref{subsec:3.3}), have been developed for this purpose.

The mean-field character of the above mentioned results is manifest in the existence of metastable states whose lifetime is infinite (in the thermodynamic limit). In realistic models (called ``finite-range'', ``finite-dimensional''  in the standard terminology used in this context), metastability is destroyed by nucleation events. Ergodicity is then restored by thermally activated processes (in the absence of a genuine thermodynamic phase transition) and  the dynamical transition at $T_c$ is smeared out, as already alluded to in the section on the mode-coupling theory.

The activated relaxation mechanisms that take over must be described in a nonperturbative way, and this represents the second stage of the theory. Kirkpatrick, Thirumalai, and Wolynes \protect\shortcite{KTW89} have proposed a description of the liquid below the crossover at $T_c$ as a ``mosaic state'' and a dynamical scaling theory close to $T_K$ based on ``entropic droplets'', the driving  force for  nucleation being provided  by  the nonzero  ``configurational entropy'' (associated with the number of free-energy minima, as found at the mean-field level). Super-Arrhenius temperature dependence of the $\alpha$ relaxation  time follows, with an effective activation barrier given in terms of the length $\xi_*$ characterizing the mosaic cells and the entropic droplets by
\begin{equation}
E(T) \sim \Delta_0 \, \xi_*(T)^{\psi},
\end{equation}
with
\begin{equation}
 \xi_*(T) \sim \left( \frac{\sigma_0}{T s_c(T)} \right)^{\frac{1}{d-\theta}},
\end{equation}
where $\Delta_0$ and $\sigma_0$ are two ``bare'' energy scales (in appropriate units), $d$ the dimension of space, and  $\psi$ and $\theta$ are two critical exponents. The latter are predicted to be both equal to $3/2$ in $d=3$ by Kirkpatrick \textit{et al.} \protect\shortcite{KTW89}, which leads to an Adam-Gibbs-type of formula for the relaxation time (see Eq.~(\ref{adam-gibbs})) with the configurational entropy per particle $s_c(T)$ vanishing at the ideal glass transition $T_K$. More recently, the mosaic scenario has been reformulated in a way that makes a direct connection between the mosaic length $\xi_*$ and a point-to-set correlation length \protect\shortcite{BBdroplet04,BB-RFOT09}, which therefore allows for possible testing \protect\shortcite{silvio_kac07,pt-to-set07,pt-to-set08,cavagnareview09}.

The last stage of the approach is formed by phenomenological input and additional modeling that are used to make contact with a broad range of experimental data in glass-forming liquids and polymers, and in glasses as well \protect\shortcite{xia00,wolynesreview07}.

The random first-order transition approach starts with a sophisticated mean-field theory which is both robust and appealing. Going beyong this and addressing nonperturbative effects is a formidable task. However, what makes a firm derivation crucial in the present case is that the mean-field scenario of a complex free-energy landscape could be very fragile to the introduction of fluctuations arising in finite range, finite dimensional systems. It has been for instance argued that the whole scenario is then destroyed \protect\shortcite{moore06}, and there is so far little evidence that it indeed persists in finite dimensions.

\subsection{Frustration-based approach: an avoided thermodynamic critical point at $T^* > T_g$}
\label{subsec:5.3}

The concept of  \textit{frustration} quite generally describes situations in which one cannot minimize the energy function of a system by merely minimizing all local interactions \protect\shortcite{toulouse}. In the context of liquids and glasses, frustration is attributed to a competition between a short-range tendency for the extension of a locally preferred order and global constraints that forbid the periodic tiling of the whole space with the local structure \protect\shortcite{sadoc-mosseri,nelson02,tarjus05}. A prototypical example is that of local tetrahedral order in three-dimensional one-component liquids in which the atoms interact through spherically symmetric pair potentials: despite being more favorable locally, extended tetrahedral or icosahedral order is precluded at large distances and cannot give rise to long-range crystalline order.
	
Frustration has first been invoked by Frank \protect\shortcite{frank52} to explain at a geometric, structural level the resistance to crystallization and degree of supercooling of a liquid, an explanation which has since received direct experimental \protect\shortcite{kelton} and numerical \protect\shortcite{charbonneau_b,charbonneau_a} confirmation. It has subsequently been used to describe the structure of glasses, more specifically metallic glasses and network forming systems \protect\shortcite{kleman79,nelson83_a,nelson83_b,nelson_etal_a,nelson_etal_b,sadoc84}. More recently, a step toward the formulation of a frustration-based theory of the glass transition has been to realize that, under rather generic conditions, frustration gives rise to the phenomenon of  ``avoided criticality'' \protect\shortcite{kivelson95,avoided_chayes96,avoided_zohar99,avoided_zohar04,tarjus05}. The latter expresses the fact that the ordering transition that may exist in the absence of frustration disappears as soon as an infinitesimal amount of frustration is introduced.

A frustration-based theory of the glass transition has been put forward, based on three plausible, but not fully established, propositions \protect\shortcite{kivelson95,more_kivelson_a,more_kivelson_b,tarjus05}: (i) the existence in a liquid of a locally preferred structure, an arrangement of molecules that minimizes some local free energy, (ii) the impossiblity for this local order to tile the whole space, which expresses ubiquitous frustration, and (iii) the possibility to construct an abstract system in which the effect of frustration can be turned off, \textit{e.g.} by tinkering with the metric of space. The resulting phenomenon of frustration-induced avoided criticality then naturally leads to collective behavior on a mesoscopic scale. In physical terms, the spatial extension of the locally preferred structure generates superextensive strain which prevents long-range ordering; below a crossover temperature $T^*$ corresponding to the transition in the absence of frustration, this results in the breaking up of the liquid into domains, whose size and further growth are limited by frustration. This occurs provided the ordering transition in the unfrustrated space is accompanied by a diverging, or at least large, correlation length and provided  frustration in a given liquid is weak enough that the transition is only narrowly avoided in physical space. 

In this frustration-limited domain picture \protect\shortcite{kivelson95,more_kivelson_a,more_kivelson_b,tarjus05}, $T^*$ marks the onset of anomalous, supermolecular behavior and it can be used to organize a scaling description of the viscous slowing down and other collective properties of glassforming liquids. For instance, the $\alpha$ relaxation time and the viscosity are predicted to follow a superArrhenius activated temperature dependence below the crossover $T^*$, the effective activation energy being expressed as $E(T)=E_{\infty}+\Delta E(T)$ with
\begin{equation}
\Delta E(T) = B T^*\left(\frac{T^*- T}{T^*}\right)^{\psi}
\end{equation}
for $T<T^*$ and the exponent $\psi$ argued from statistical-mechanical and phenomenological arguments to be close to $8/3$ in $d=3$. The ``fragility'' of a glassformer, \textit{i.e.} the  departure from Arrhenius behavior, is quantified by the parameter $B$ which is inversely proportional to the degree of frustration. In this approach, a large fragility is therefore associated with a small frustration, which implies a closer proximity to the avoided transition and a larger extent of collective behavior. Moreover, as in the random first-order transition theory, the heterogeneity of the dynamics primarily stems from the ``patchwork'' or ``mosaic'' character of the configurations.

The frustration-based theory leads to the formulation of effective minimal models of glass-forming systems for which genuine statistical-mechanical treatments and numerical computations are possible \protect\shortcite{grousson_a,grousson_b,sausset08,sausset10}. It faces however two main difficulties. A first one  is technical, and it is shared by all other theories of relaxation in deeply supercooled liquids: the phenomena associated with avoided criticality are intrinsically nonperturbative and a full-blown resolution going beyond phenomenological modeling and limited computations is \textit{a priori} very hard. The second one concerns the very physical basis of the approach: the ubiquitousness of frustration in glass-forming liquids is a resonable postulate but it requires confirmation. The icosahedral example has presumably no value in molecular liquids and polymers, and one still awaits proper identification of a locally preferred structure for molecules of nonspherical shapes that form most real fragile glassformers.

As a final comment, I would like to point out that an appealing property of the frustration approach, whether or nor it represents the ``general theory'' of the glass transition, is that it produces microscopic models in which the fragility of a glass-former, and more generally the degree to which collective behavior can develop, can be varied at will by tuning the amount of frustration, while keeping the other parameters such as the interaction potentials fixed \protect\shortcite{sausset08}.

\subsection{Jamming scenario: an unreachable static critical point at $T=0$}
\label{subsec:5.4}

The \textit{jamming} scenario and the associated phase diagram may be viewed as a grand unification scheme in which the glass transition of liquids and polymers is taken as one example of a more general phenomenon in which the sluggish response of a condensed-matter system leads to an amorphous arrested state with no observable macroscopic flow \protect\shortcite{jam-diag98,jam_book}. At the heart of this approach is the realization that temperature, packing fraction and stress act similarly on a disordered system on the verge of rigidity. This allows one to draw analogies between granular materials, emulsions and foams under the effect of external stress or forcing, repulsive colloidal suspensions as a function of concentration, and glass-forming liquids and polymers with temperature as the control variable \protect\shortcite{jamming,jam_book}. 

The jamming paradigm may prove useful in suggesting systematic experimental investigations of a given material as a function of several parameters potentially controlling its jamming or unjammming behavior:  for instance glass formation and the associated phenomena could be studied in liquids and polymers not only through temperature or pressure changes but also by varying the applied stress. However, for interesting as they may be, a heuristic phase diagram and broad-based comparisons do not make a theory. A step forward in establishing a jamming-based theory has been the evidence for the existence of a well-identified, genuine, critical point located at zero temperature \protect\shortcite{Ohern02_a,Ohern02_b,bootstrap}.

``Point J'', the jamming critical point, is observed in model systems of spherical particles interacting through finite-range repulsive potentials as one compresses the system via a nonequilibrium protocol. Scaling laws and a diverging (static) correlation length are found around point J \protect\shortcite{Ohern02_a,Ohern02_b,bootstrap}, and the marginally rigid jammed solid close to point J shows an anomalous elastic response characterized by the presence of soft (zero or low-frequency) vibrational modes \protect\shortcite{wyart05,softmodes,jam_book}.  Point J is hypothesized to control the jamming behavior of soft-condensed systems as foams and emulsion, of hard objects such as solid colloidal particles and grains  (point J could then be interpreted as some sort of random close packing at which the pressure first diverges), as well as glass-forming liquids and polymers.

Several issues have been raised concerning the jamming scenario and its associated zero-temperature critical point:

(i) The robustness of point J with respect to physically relevant factors not included in the original formulation: friction for grains, asphericity of particle shape, thermal fluctuations, longer-ranged interaction potentials and attractive forces; some of these factors can be accounted for by an appropriate generalization \protect\shortcite{jam_book,van_hecke}, but the latter one seems more problematic \protect\shortcite{ludo-gilles09}.

(ii) The uniqueness of point J. The precise location of point J is protocol dependent and actually a whole range of J-points seems to exist at zero temperature \protect\shortcite{protocol07,protocol09}. As itself, this is not a fatal blow to the jamming picture, but it requires some caution.

(iii) The relevance of point J to the glass transition of liquids and polymers, which is the main focus of this overview. The interactions among particles in liquids and polymers involve long-range attractive interactions. As already mentioned, treating the latter as a mere perturbation is highly questionable \protect\shortcite{ludo-gilles09}.

Whereas some specific phenomena associated with slow dynamics in liquids and anomalous elastic response in glasses may profitably be envisaged within the jamming context, it is at present unclear if  the physics of glass formation can be reduced to a jamming behavior controlled by a zero-temperature  point J.

\section{Concluding remarks}
\label{sec:6}

How to assess the validity of the proposed theories? This seemingly trivial question is more complex than it may sound at first. Indeed, theories mostly rationalize existing data. In addition to providing a narrative to explain the qualitative trends that characterize glass formation, they reproduce quantitative features of the phenomenology, but at the expense of adjustable parameters. New predictions that would not involve additional assumptions and could be crisply checked in experiments are rare if not inexistent.

Experimental observations do put constraints on the theories. Phenomena, trends, orders of magnitude and correlations found in glass-forming systems should of course be reproduced. However, there is always some leeway. First, some of the observations may be discarded by the proponents of a theory as irrelevant or out of the scope of their approach.  Second, many of the theories, especially those attempting to describe the viscous regime close to the glass transition, have phenomenological input and  cannot just be taken as ``first-principle'' approaches. (Even the mode-coupling theory, which is often considered as the archetype of a ``microscopic'' theory and certainly makes detailed predictions based on liquid structure factors, cannot be tested without bias because of the loosely defined, and therefore adjustable, location of the crossover replacing the dynamical singularity $T_c$ in real systems.) The resulting unavoidable use of fitting parameters allows for flexibility in  comparing theory with experiment. Not surprisingly, several theories can claim success in reproducing observed properties, for instance the temperature dependence of the relaxation time and the viscosity, in spite of the quite different functional forms and underlying physics they entail.

The above considerations also apply to comparisons with data obtained through computer simulation of realistic models of liquids. Simulations may  nonetheless offer a possibility of testing the theoretical premises of the proposed approaches in some depth. Such studies have already been undertaken on basic issues such as the existence of growing point-to-set correlations and of a nonzero surface tension between amorphous glassy metastable states, evidence for locally preferred structures in atomic and molecular liquids and relevance of frustration, robustness of the critical jamming point J to the introduction of friction or of long-ranged attractive forces, the appearance of slow anomalous modes associated with marginal rigidity, etc.

Constraints on theories also lie in their internal consistency. However, it may be hard to maintain full rigor when tackling ``nonperturbative'' effects which, in my opinion, are at the core of the glass transition phenomenon (see above). More likely,  one will at best be able to study limiting cases and possibly to make asymptotic predictions when the postulated collective behavior is dominant. In this endeavor, as in computer simulations, progress seems to be limited by the absence of a widely accepted and sufficiently tractable minimal model. A reasonable strategy to assess the validity of the candidates for a ``general theory'' of the glass transition would then be to devise models in which the collective behavior is so exagerated that asymptotic predictions can be cleanly proved or disproved, without having to worry about subdominant  effects \protect\shortcite{kiv-tar08}. Such models should of course allow one to go continuously, with quantitative but no qualitative changes, from this extreme, but testable, behavior to that of real glass-forming systems.

To conclude this overview of theories of the glass transition, I would like to raise an apparently incongruous question: to which extent could various theoretical approaches be compatible or even complementary? The scenarii and narratives of glass formation produced by most theories do seem completely at odds. However, this may be less so for the developed frameworks, the working concepts and the proposed relaxation mechanisms that underly these theories. To illustrate this with a few examples: frustration is undoubtedly compatible with the emergence of a complex free-energy landscape and with a description of supercooled liquids as mosaic states, and none of these theoretical elements are in contradiction with the development of kinetic constraints on the motion of some effective degrees of freedom and with the presence of facilitation; in a somewhat different vein, elastic models could be extended and combined with the effect of a growing length scale as the glass transition is approached. Going further in a fruitful way of course requires a serious elaboration to avoid the risk of diluting even more the assessment of the theoretical constructions. However, at this stage where, despite the progress made in the last $25$ years, full resolution of the glass transition problem does not seen around the corner, it may be wise to take an open view on the possible complementarity of different theoretical perspectives.

\bibliographystyle{OUPnamed_notitle}
\bibliography{refs_chap2raw}

\end{document}